\PassOptionsToPackage{dvipsnames}{xcolor}
\PassOptionsToPackage{colorlinks,citecolor=Blue,linkcolor=Red,urlcolor=Cerulean}{hyperref}
\documentclass{SciPost}
\usepackage{amsmath}
\usepackage{amssymb}
\usepackage{graphicx}
\usepackage{xcolor}

\global\long\def\E{\mathrm{e}}

\global\long\def\I{\mathrm{i}}

 \global\long\def\ket#1{\left|#1\right\rangle }

 \global\long\def\bra#1{\left\langle #1\right|}

\begin{document}

\begin{center}{\Large \textbf{
Absence of dynamical localization in interacting driven systems
}}\end{center}

\begin{center}
David J. Luitz\textsuperscript{1*}, 
Yevgeny Bar Lev\textsuperscript{2}, 
Achilleas Lazarides\textsuperscript{3}
\end{center}

\begin{center}
\textbf{1} {Department of Physics, T42, Technische Universität München,\\ James-Franck-Straße
1, D-85748 Garching, Germany}

\textbf{2} {Department of Chemistry, Columbia University,\\ 3000 Broadway, New
York, New York 10027, USA}

\textbf{3} {Max-Planck-Institut für Physik komplexer Systeme, 01187 Dresden,
Germany}

* {david.luitz@tum.de}
\end{center}

\begin{center}
\today
\end{center}


\section*{Abstract}
{\bf 
Using a numerically exact method we study the stability of dynamical
localization to the addition of interactions in a periodically
driven isolated quantum system which conserves only the total number of particles.
We find that while even infinitesimally small interactions destroy
dynamical localization, for weak interactions density transport is
significantly suppressed and is asymptotically diffusive, with a diffusion
coefficient proportional to the interaction strength. 
For systems tuned away from dynamical localization, transport is dramatically enhanced and diffusion
is observable only for sufficiently small detuning.
}

\vspace{10pt}
\noindent\rule{\textwidth}{1pt}
\tableofcontents\thispagestyle{fancy}
\noindent\rule{\textwidth}{1pt}
\vspace{10pt}

\section{Introduction}

The last decade has witnessed an increasing interest in isolated out-of-equilibrium
(OOE) quantum systems both in theory and experiment. In particular,
periodically-driven, or Floquet many-body systems have become one
of the most active areas in condensed matter theory and many-body
physics, due to the exciting possibilities of finding novel OOE phases
of matter, without an analog in static systems. 

Driven interacting systems typically heat up, until they approach a
nonequilibrium steady-state (NESS) ``locally identical'' to a
featureless state of maximal Gibbs entropy. That is, \emph{local}
observables cannot differentiate between the two states
\cite{DAlessio2014,Lazarides2014b,Ponte2014a}.

This ``local identity'' between the NESS and the maximal Gibbs entropy
state is broken when conservation laws exist, for example 
in noninteracting systems \cite{Russomanno2012,Lazarides2014a}. The
systems still approach a NESS, but this NESS is ``locally
identical'' to a state which maximizes the Gibbs entropy under \emph{all}
the constraints imposed by the conserved quantities. A large body of
work has concentrated on ``Floquet engineering'' of such
noninteracting systems, namely using periodic driving as a tool to
create effective time-independent Hamiltonians of desirable but
difficult to implement static Hamiltonians
\cite{Holthaus:2015ka,Bukov2015,Eckardt2017}.

For interacting systems it is possible to avoid the featureless NESS
by the addition of a quenched disorder.  In the
absence of driving, these systems are many-body localized (MBL)
\cite{Basko2006a,Agarwal2016_review,Nandkishore2014,Luitz2016c,Abanin2017,Imbrie2016a,Lazarides2014,Ponte2014},
and for sufficiently high driving frequencies they do not heat
up indefinitely \cite{Lazarides2014,Ponte2014}. Such systems were also
shown to host nontrivial eigenstate order, corresponding to a new OOE
phase of matter, colloquially dubbed a
``time-crystal''\cite{Khemani2015a} and subsequently studied in both
experimental and theoretical
studies~\cite{Else:2016ue,Yao2016,Zhang2016c,Choi2016a}.

In the present work we explore a different potential avenue to suppress the indefinite
heating of interacting driven systems by starting with a noninteracting
translationally invariant system which is dynamically localized \cite{Revie1986,Dunlap1988,Grossmann1991,Eckardt:2009gz,Das2010}
(see also the closely related study of a \emph{disordered} system
\cite{Bairey2017}, which appeared while our work was in preparation).
The special form of the driving term in dynamically localized systems
leads to a frozen stroboscopic dynamics and therefore to disorder-free
spatial localization. Whether adding interactions to such disorder-free
localized systems results in heating, and what is the nature of the
transport are the main questions we consider in our work. We would
like to stress that the mechanism behind this localization, while
of quantum nature, is \emph{not} due to Anderson localization, as
happens in related disorder-free kicked rotor systems which unfortunately
are also dubbed ``dynamically localized'' \cite{Fishman1982,Grempel1982,Grempel1984}.
The interacting versions of such systems were recently studied, but
are not of direct relevance to our work \cite{Keser2016,Rozenbaum2017}. 

The structure of this work is as follows: We present the model in
Section \ref{sec:Model}, briefly survey dynamical localization adapted
to the many-body context in Section \ref{sec:Dynamical-localization}
and in Section \ref{sec:Results} present our main results showing
that interactions destroy dynamical localization and cause diffusive
dynamics with a diffusion coefficient proportional to the interaction.
We end with our conclusions, a brief discussion of the implications
of our results for the locality of Floquet effective Hamiltonians
and an outlook.

\section{\label{sec:Model}Model}

\begin{figure}
\begin{centering}
    \includegraphics[width=\textwidth]{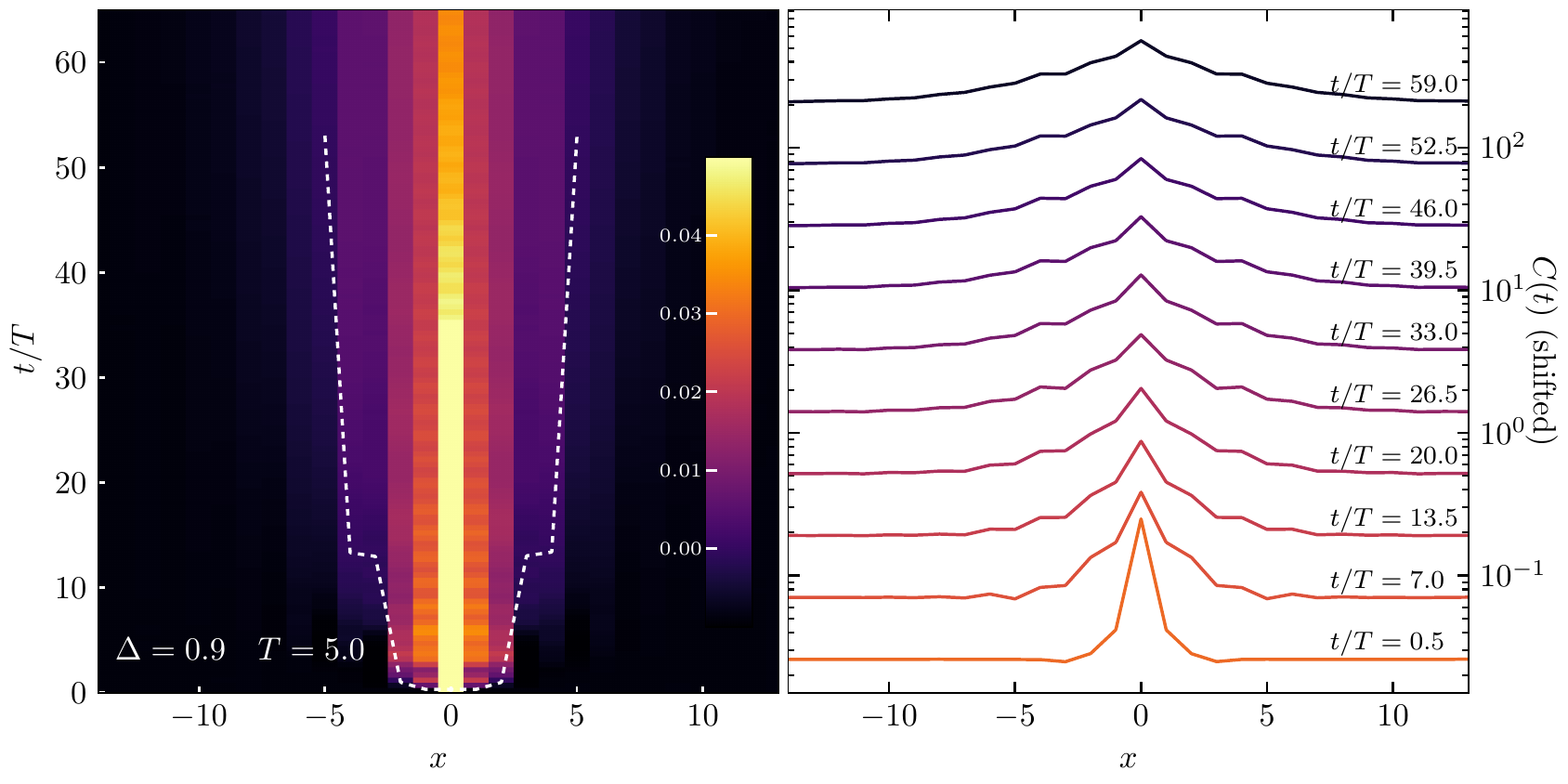}
\par\end{centering}
\caption{\label{fig:Spreading-of-density}Spreading of a density excitation.
The left panel shows the contour plot of the excitation quantified by $C_x(t)$ as a function of time
$t$ and space $x$ (cf. Eq. \eqref{eq:density-dnesity-corr}), with more
intense colors representing larger values. The white line is an isoline
set at $C_{x}\left(t\right)=0$. The right panel shows cuts through
the excitation profile at different times (origin offset for readability). The
parameters used in this simulation are: $L=29$, $T=5$, $\Delta=0.9$
and the amplitude of the drive is set to the dynamical localization
point $A_{0}=4\pi/T$ (\ref{eq:dynloc_pnt}).}
\end{figure}
In this work we consider a driven Hamiltonian of the form,
\begin{equation}
\hat{H}=\hat{H}_{0}+f\left(t\right)\sum_{m=1}^{L}m\hat{n}_{m},\label{eq:driven_heisenberg}
\end{equation}
where the static Hamiltonian is given by
\begin{equation}
\hat{H}_{0}=-\frac{J}{2}\sum_{m=1}^{L-1}\left(\hat{c}_{m}^{\dagger}\hat{c}_{m+1}+\hat{c}_{m+1}^{\dagger}\hat{c}_{m}\right)+\Delta\sum_{m=1}^{L-1}\hat{n}_{m}\hat{n}_{m+1}.\label{eq:interacting-ham}
\end{equation}
Here $L$ is the length of the lattice, $\hat{c}_{m}^{\dagger}$ creates
a spinless fermion at site $m$, $\hat{n}_{m}$ is the number operator,
$J$ is the hopping and $U$ is the interaction strength, and the
driving protocol is
\begin{equation}
f\left(t\right)=\begin{cases}
-A & -\frac{T}{2}\leq t<0\quad\left(\mod\,T\right)\\
A & 0\leq t<\frac{T}{2}\quad\left(\mod\,T\right),
\end{cases}\label{eq:driveterm}
\end{equation}
where $T=2\pi/\omega$ is the period of the driving, $\omega$ is
the driving frequency and $A$ is the amplitude of the drive. This
model corresponds to a constant potential gradient with a strength proportional
to $A$, the sign of which is flipped every half period.

\section{\label{sec:Dynamical-localization}Dynamical localization}

For $\Delta=0$, the model (\ref{eq:driven_heisenberg}) is noninteracting
and exactly solvable \cite{Revie1986,Dunlap1988}. For specific ratios
of $A/\omega$ it exhibits dynamical localization, namely, an initially
localized density excitation \emph{does not spread over time}. For
convenience of the reader in this Section we adapt the derivation
of dynamical localization of Ref.~\cite{Dunlap1988} to the many-body
setting. For this purpose it is instructive to perform a \emph{time-dependent}
unitary transformation which eliminates the drive. This can be always
achieved, since setting $\ket{\psi}=V\left(t\right)\ket{\phi}$ and
using the Schrödinger equation, $\I\partial_{t}\ket{\psi}=\hat{H}\ket{\psi}$
it is easy to show that $\ket{\phi}$ satisfies $\I\partial_{t}\ket{\phi}=\tilde{H}\left(t\right)\ket{\phi},$
with the transformed Hamiltonian,
\begin{equation}
\tilde{H}\left(t\right)=V^{\dagger}\left(t\right)H\left(t\right)V\left(t\right)-\I V^{\dagger}\left(t\right)\partial_{t}V\left(t\right).
\end{equation}
Therefore if the original Hamiltonian is of the following form,
\begin{equation}
\hat{H}=\hat{H}_{0}+f\left(t\right)\hat{H}_{1},
\end{equation}
the second term can be eliminated by choosing,
\begin{equation}
\hat{V}=\E^{-\I\mathcal{A}\left(t\right)\hat{H}_{1}},
\end{equation}
where $\mathcal{A}\left(t\right)=\int_{0}^{t}\mathrm{d}\bar{t}\,f\left(\bar{t}\right).$
The transformed Hamiltonian is then given by,
\begin{equation}
\tilde{H}=\E^{\I\mathcal{A}\left(t\right)\hat{H}_{1}}\hat{H}_{0}\E^{-\I\mathcal{A}\left(t\right)\hat{H}_{1}}.
\end{equation}
While the original driving term was eliminated, it appears that nothing
was gained since the transformed Hamiltonian is still time-dependent.
For a spatially uniform and temporally periodic force on the system,
namely,
\begin{equation}
\hat{H}_{1}=\sum_{k}k\hat{n}_{k},
\end{equation}
further progress can be made. For such a drive the destruction and
creation operators transform as,

\begin{equation}
\hat{a}_{m}=\E^{-\I\mathcal{A}\left(t\right)m}\hat{c}_{m}\qquad\hat{a}_{m}^{\dagger}=\E^{\I\mathcal{A}\left(t\right)m}\hat{c}_{m}^{\dagger},
\end{equation}
and the number operators are invariant under this transformation,
$\hat{n}_{m}=\hat{c}_{m}^{\dagger}\hat{c}_{m}=\hat{a}_{m}^{\dagger}\hat{a}_{m}$.
Therefore a generic Hamiltonian of the form,
\begin{equation}
\hat{H}_{gen}=\sum_{n\neq m}h_{nm}\hat{c}_{n}^{\dagger}\hat{c}_{m}+g\left(\left\{ \hat{n}_{i}\right\} \right)+f\left(t\right)\sum_{l}l\hat{n}_{l},
\end{equation}
will be transformed to,

\begin{equation}
\tilde{H}=\sum_{n\neq m}\E^{\I\mathcal{A}\left(t\right)\left(n-m\right)}h_{nm}\hat{a}_{n}^{\dagger}\hat{a}_{m}+g\left(\left\{ \hat{n}_{i}\right\} \right),\label{eq:transformed_ham}
\end{equation}
with time-dependence appearing only in the hopping term and with an
arbitrary function $g(\{\hat{n}_{i}\})$ depending only on the number
operators. For simplicity we will now restrict the discussion to a
single band by setting $h_{nm}=\frac{J}{2}\left(\delta_{n,m-1}+\delta_{n,m+1}\right)$.
The total current is given by
\begin{equation}
\hat{J}=-\I\frac{J}{2}\sum_{n}\left(\E^{\I\mathcal{A}\left(t\right)}\hat{a}_{n}^{\dagger}\hat{a}_{n+1}-\E^{-\I\mathcal{A}\left(t\right)}\hat{a}_{n+1}^{\dagger}\hat{a}_{n}\right),
\end{equation}
which can be diagonalized to
\begin{equation}
\hat{J}=-J\sum_{k}\sin\left(k+\mathcal{A}\left(t\right)\right)\hat{f}_{k}^{\dagger}\hat{f}_{k},\label{eq:total_current_diag}
\end{equation}
where $\hat{f}_{k}=\sum_{n}\exp\left(\I kn\right)\hat{a}_{n}/\sqrt{L}$.
For $g\left(\left\{ \hat{n}_{i}\right\} \right)=0$, namely in the
absence of interactions and external potentials, the Hamiltonian can
be diagonalized simultaneously with $\hat{J}$, and therefore also
the transformed Floquet operator,
\begin{equation}
\hat{U}\left(T,0\right)\equiv\E^{-\I\int_{0}^{T}\mathrm{d}\bar{t}\tilde{H}\left(\bar{t}\right)}=\E^{-\I J\sum_{k}\hat{f}_{k}^{\dagger}\hat{f}_{k}\int_{0}^{T}\mathrm{d}\bar{t}\cos\left(k+\mathcal{A}\left(\bar{t}\right)\right)}\label{eq:Floquet_op}
\end{equation}
can be computed. Although the current is \emph{not} conserved its
expectation can be readily calculated,
\begin{equation}
\left\langle \hat{J}\right\rangle =-J\sum_{k}\sin\left(k+\mathcal{A}\left(t\right)\right)\left\langle \hat{f}_{k}^{\dagger}\hat{f}_{k}\right\rangle .
\end{equation}
Dynamical localization occurs when the total number of particles transported
in one period vanishes, that is, when the integral of the current
over one period yields zero for all $k$, 
\begin{equation}
\int_{-T/2}^{T/2}\mathrm{d}t\E^{\I\mathcal{A}\left(t\right)}=0.
\end{equation}
In this case the Floquet operator (\ref{eq:Floquet_op}) reduces
to the identity and its spectrum collapses such that all its eigenvalues
are degenerate, making the stroboscopic dynamics trivial. Concretely,
for the driving we use in this work (\ref{eq:driveterm}),
$\mathcal{A}\left(t\right)=A\left|t\right|$ (for $\left|t\right|<T/2$),
and we have,
\begin{equation}
\int_{-T/2}^{T/2}\mathrm{d}t\E^{\I\mathcal{A}\left|t\right|}=2\int_{0}^{T/2}\mathrm{d}t\E^{\I At}=2\frac{\E^{\I AT/2}-1}{\I A},
\end{equation}
which vanishes for $\E^{\I AT/2}=1$, or $A/\omega=2n\quad n\in\mathbb{Z}.$ 

When an external potential or interactions are present the Hamiltonian
cannot be diagonalized in the basis of $\hat{J}$ and the above derivation
does not apply since quasimomentum is not conserved. Notwithstanding,
one might still expect some residual suppression of transport, since
using the periodicity of $\exp[\I\mathcal{A}\left(t\right)]=\sum_{n}C_{n}\exp\left[\I\omega nt\right]$
one can cast the generic Hamiltonian (\ref{eq:transformed_ham}) into
the form
\begin{equation}
\tilde{H}=\sum_{n\neq m}C_{0}h_{nm}\hat{c}_{n}^{\dagger}\hat{c}_{m}+g\left(\left\{ \hat{n}_{i}\right\} \right)+\sum_{n\neq m}h_{nm}\hat{c}_{n}^{\dagger}\hat{c}_{m}\sum_{s}C_{s}e^{i\omega st},\label{eq:transformed_ham_expanded}
\end{equation}
which includes a static Hamiltonian with a hopping term renormalized
by $C_{0}$. This description is however oversimplified, due
to the presence of a nontrivial driving term, which is coupled to the hopping. 
In the following Section we will explore what happens to transport
in an interacting model, where the first term is set to vanish, namely
$C_{0}=0$. 

\section{\label{sec:Results}Results}

\begin{figure}[h]
\begin{centering}
    \includegraphics[width=\textwidth]{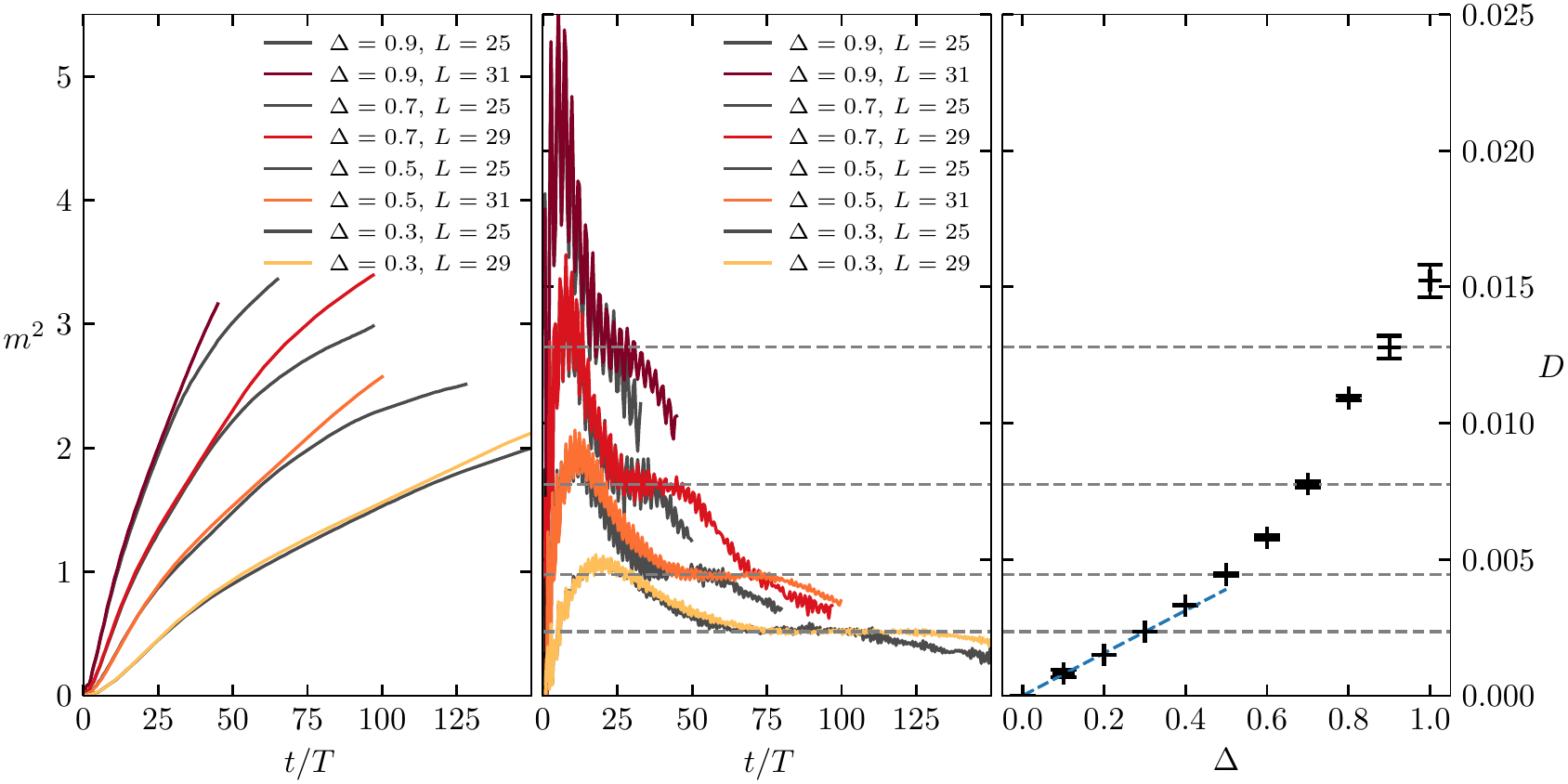}
\par\end{centering}
\caption{\label{fig:MSD-deriv}Mean-square displacement $m^2$ (cf. Eq. \eqref{eq:msd}) of the density excitation
(left) and its derivative, the time-dependent diffusion coefficient
$D\left(t\right)$, (middle) as a function of time for a few interaction
strengths, $\Delta=0.3,0.6$ and $0.9$ (stronger interaction is indicated
by more intense color). The colored lines indicate system sizes $L=29$
and $L=31$ and the grey lines correspond to $L=25$, simulated
with the same parameters. The horizontal dashed lines demonstrate
the extraction of the diffusion coefficient from the flat part of
the derivative, and the right panel shows the diffusion coefficient
as a function of the interaction strength. The parameters used in
this simulation are: $T=5$ and the amplitude of the drive is set
to the dynamical localization point $A_{0}=4\pi/T$ (\ref{eq:dynloc_pnt}).}
\end{figure}

We study density transport in a driven and interacting system. For
this purpose we calculate the density-density correlation function,
\begin{equation}
C_{x}\left(t\right)=\frac{1}{Z}\text{Tr }\left[\left(\hat{n}_{x}\left(t\right)-\frac{1}{2}\right)\left(\hat{n}_{0}-\frac{1}{2}\right)\right],\label{eq:density-dnesity-corr}
\end{equation}
which describes the spreading of density excitation in the system.
Here $Z$ is the dimension of the relevant Hilbert space, and the
operators are written in the Heisenberg picture with respect to the
\emph{driven} Hamiltonian. Intuitively this describes the spreading
of a density excitation in the presence of driving. Here, we fix the number of
fermions to $N=\left(L-1\right)/2$, and use odd sizes $L$, such
that an excitation at the center is at the same distance from the left and right
(open) boundaries. To compute the excitation profile we stroboscopically
evolve two initial states over $n$ periods, $\ket{\psi_{R}\left(n\right)}=\hat{U}\left(nT\right)\left(\hat{n}_{0}-\frac{1}{2}\right)\ket{\psi_{0}}$
and $\ket{\psi_{L}\left(n\right)}=\hat{U}\left(nT\right)\ket{\psi_{0}}$,
using a numerically exact method. The method is based on projecting the time evolution operator 
to an orthonormal basis of a truncated Krylov space spanned by the initial state, which is updated at
each time step (for a pedagogical review of the method, see Sec. 5.1.2 of Ref. \cite{Luitz2016c}).
We then calculate
the matrix element $C_{x}\left(nT\right)=\left\langle \psi_{L}\left(n\right)\left|\left(\hat{n}_{x}-\frac{1}{2}\right)\right|\psi_{R}\left(n\right)\right\rangle $.
For the driving protocol we use in this work (\ref{eq:driveterm})
the propagator $\hat{U}\left(nT\right)$ over $n$ periods is simplified
to,
\begin{equation}
    \hat{U}\left(nT\right)\equiv \mathcal{T} \exp\left[-\I\int_{0}^{nT}\hat{H}\left(\bar{t}\right)\mathrm{d}\bar{t}\right]=\prod_{i=1}^{n}\left(\E^{-\I\hat{H}_{-}T/2}\E^{-\I\hat{H}_{+}T/2}\right),\label{eq:definition-U}
\end{equation}
where $\hat{H}_{\pm}$ corresponds to the driven Hamiltonian (\ref{eq:driven_heisenberg})
in the first and second halves of the period and $\mathcal{T}$ is the time ordering operator. The initial state $\ket{\psi_{0}}$
is sampled randomly from the Haar measure, which due to Lévy's Lemma
approximates the trace in (\ref{eq:density-dnesity-corr}) with an
error which is exponentially small in the system size \cite{Ledoux2005,Luitz2016c}.
This computational scheme allows us to reach systems sizes as large as $L=31$ sites, which
corresponds to a Hilbert space dimension of $300,540,195$ (see Ref.~\cite{Luitz2016c}
for more details on the numerical procedure).

\begin{figure}[h]
\begin{centering}
    \includegraphics[width=\textwidth]{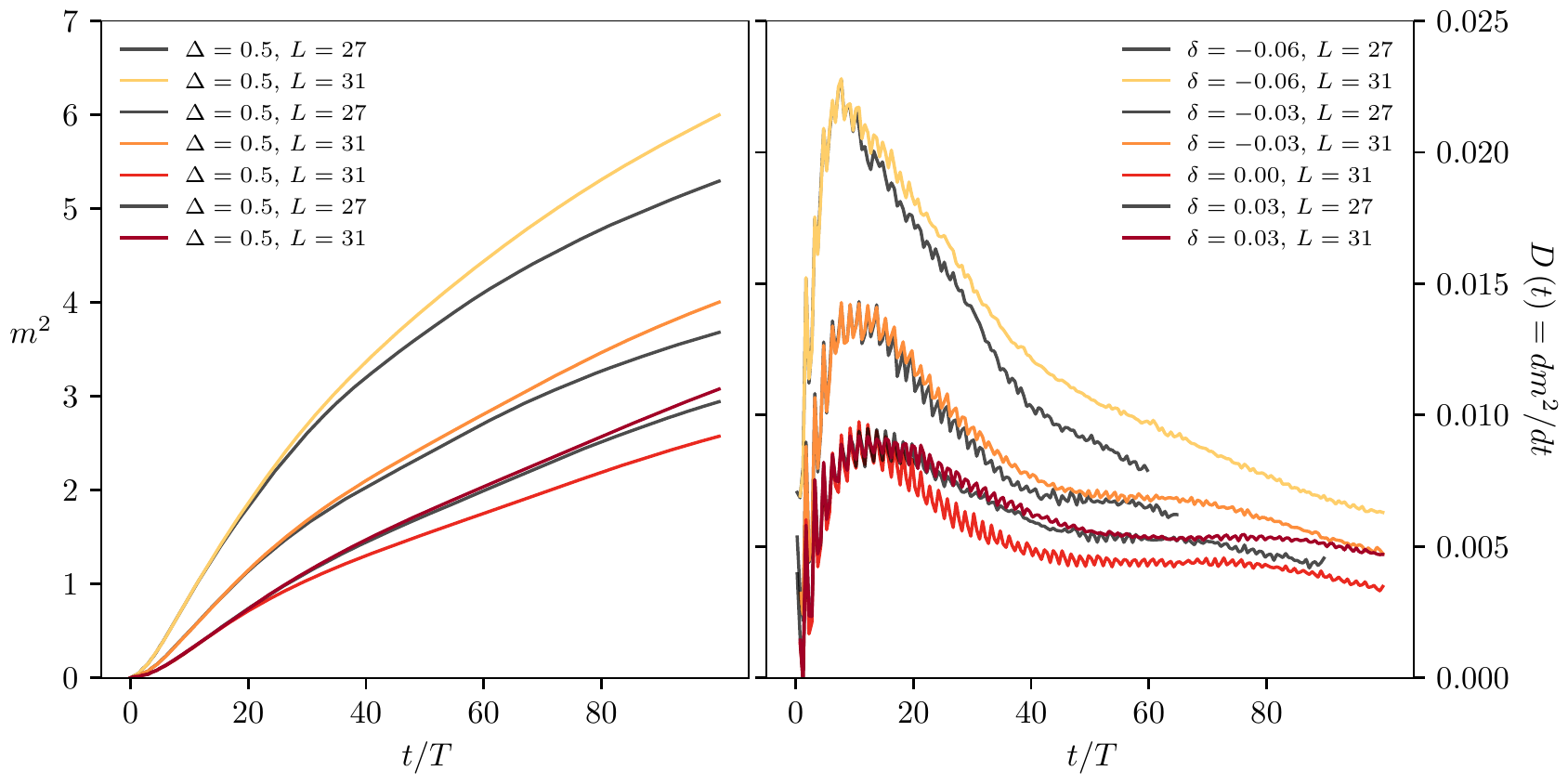}
\par\end{centering}
\caption{\label{fig:MSD-deriv-detun}Same as Fig.~\ref{fig:MSD-deriv}, but
detuning from the dynamical localization point by changing the amplitude of the drive, $\delta=0, \pm0.03$
    and $-0.06$ (see (\ref{eq:detuning-definition})). The colored lines indicate
system size $L=31$ and the grey lines systems size $L=27$, simulated
with the same parameters. The interaction strength is fixed at $\Delta=0.5$.}
\end{figure}

To characterize the transport we calculate the mean-square displacement
(MSD) of the density excitation,
\begin{equation}
m^{2}\left(t\right)\equiv\sum_{x}x^{2}\left[C_{x}\left(t\right)-C_{x}\left(0\right)\right],\label{eq:msd}
\end{equation}
and the time-dependent diffusion coefficient \cite{Steinigeweg2009a,Yan2015,Luitz2016c,Steinigeweg2017},
\begin{equation}
D\left(t\right)\equiv\frac{\mathrm{d}m^{2}}{\mathrm{d}t}.\label{eq:diff_coff}
\end{equation}
We note in passing that within the linear response theory, namely
when the drive is taken to be a small perturbation to the equilibrium
state, this diffusion coefficient computed in the limit of $t\to\infty$
corresponds to the value calculated from the Kubo formula. In this
work we however \emph{do not consider the linear response regime};
our driving amplitudes are large and take the state of the system
very far from equilibrium. Therefore there is no reason to expect
that the diffusion coefficient calculated from (\ref{eq:diff_coff})
and the Kubo formula coincide. In fact it is not even clear \emph{a priori}
if transport in such case will be diffusive. The dynamical localization
of noninteracting systems, which was discussed in the previous Section,
serves as an example when the diffusion coefficient (depending on
the drive amplitude) is either zero or infinity and there is either
no transport or the transport is ballistic. 

We now proceed to examine the stability of dynamical localization
to the addition of interactions. To this end we set the system to
the noninteracting dynamical localization point by setting the amplitude
of the drive to, 
\begin{equation}
A_{0}/\omega=2\qquad A_{0}=4\pi/T.\label{eq:dynloc_pnt}
\end{equation}
Throughout this work we fix the period of the drive to be $T=5$,
such that the frequency, $\omega=2\pi/T\approx1.256$ is smaller then
the one-particle and many-body bandwidths, to allow effective energy
distribution in the system using single and many-particle rearrangements.
While the system is in the \emph{noninteracting} dynamical localization
point, the left panel of Fig.~\ref{fig:Spreading-of-density} shows
sub-ballistic spreading of the density excitation, indicating that
dynamical localization is unstable under the addition of interactions.
The right panel shows fixed time cuts through the excitation profile.
Interestingly, even after relatively long times these profiles do
not approach a Gaussian form, suggesting that the density-density
correlation function $C_{x}\left(t\right)$ is \emph{not} well described
by a diffusion equation. We also added an isoline at $C_{x}(t)=0$
(the background of $C_{x}(t)$ is slightly negative due to the trivial
correlation imposed by particle number conservation), which shows a
peculiar feature at short times. We attribute this feature to short time
transient dynamics which roughly corresponds to the location of the peak in the time dependent diffusion
coefficient $D(t)$ shown in Fig. \ref{fig:MSD-deriv}. The asymptotic
transport is only reached at later times.

To characterize the nature of transport, and to estimate the finite
size effects in the system we calculate the MSD (\ref{eq:msd}) for
different interaction strengths and several system sizes (not all
shown for readability). From the left panel of Fig.~\ref{fig:MSD-deriv}
it is clear that while dynamical localization is destroyed by the
addition of interactions, for weak interactions transport is still
significantly suppressed and signalled by a slow growth of the MSD,
which allows us to reach very long times before our results are affected
by finite size effects. This time becomes shorter for stronger interactions,
as could be inferred from the deviation in MSD between the various
system sizes. After a short time of fast transient transport the MSD
appears to enter a diffusive regime, as could be seen from a linear growth of the MSD. 
To see it more clearly
we compute the time-dependent diffusion coefficient (\ref{eq:diff_coff})
in the middle panel of Fig.~\ref{fig:MSD-deriv}. It shows a clear
saturation to finite plateaus, which become longer for larger system
sizes (while the height of the plateaus remains the same) and indicate that the
departure from the plateau at later times is a finite size effect.
The asymptotic diffusion coefficient is extracted from the height
of the plateaus for different interaction strengths and is presented
in the right panel of Fig.~\ref{fig:MSD-deriv}. While the overall
dependence on the interaction strength is nonlinear, for sufficiently
weak interaction strengths we find $D\sim\Delta$.

We now fix the interaction strength to $\Delta=0.5$, such that we
are able to reach a pronounced diffusive regime within our simulation
times for $A/A_{0}=1$, and detune away from the dynamical localization
point by slightly changing the amplitude of the drive. We characterize
the strength of the detuning from the dynamical localization point by
\begin{equation}
    \delta=\frac{A-A_{0}}{A_{0}}.\label{eq:detuning-definition}
\end{equation}
Fig.~\ref{fig:MSD-deriv-detun} shows the results for three different detuning strengths and two system sizes. 
Transport is much faster in this regime even for a few percent detuning from
the dynamical localization point, requiring very large system sizes
to capture the asymptotic transport. Both the MSD and the
time-dependent diffusion coefficient show diffusive behavior for small
detuning $\delta=\pm0.03$, while for larger (in magnitude) detuning $\delta=-0.06$
it appears that our results do not capture the asymptotic transport
even for the largest system size $L=31.$

\begin{figure}
\begin{centering}
    \includegraphics[width=\textwidth]{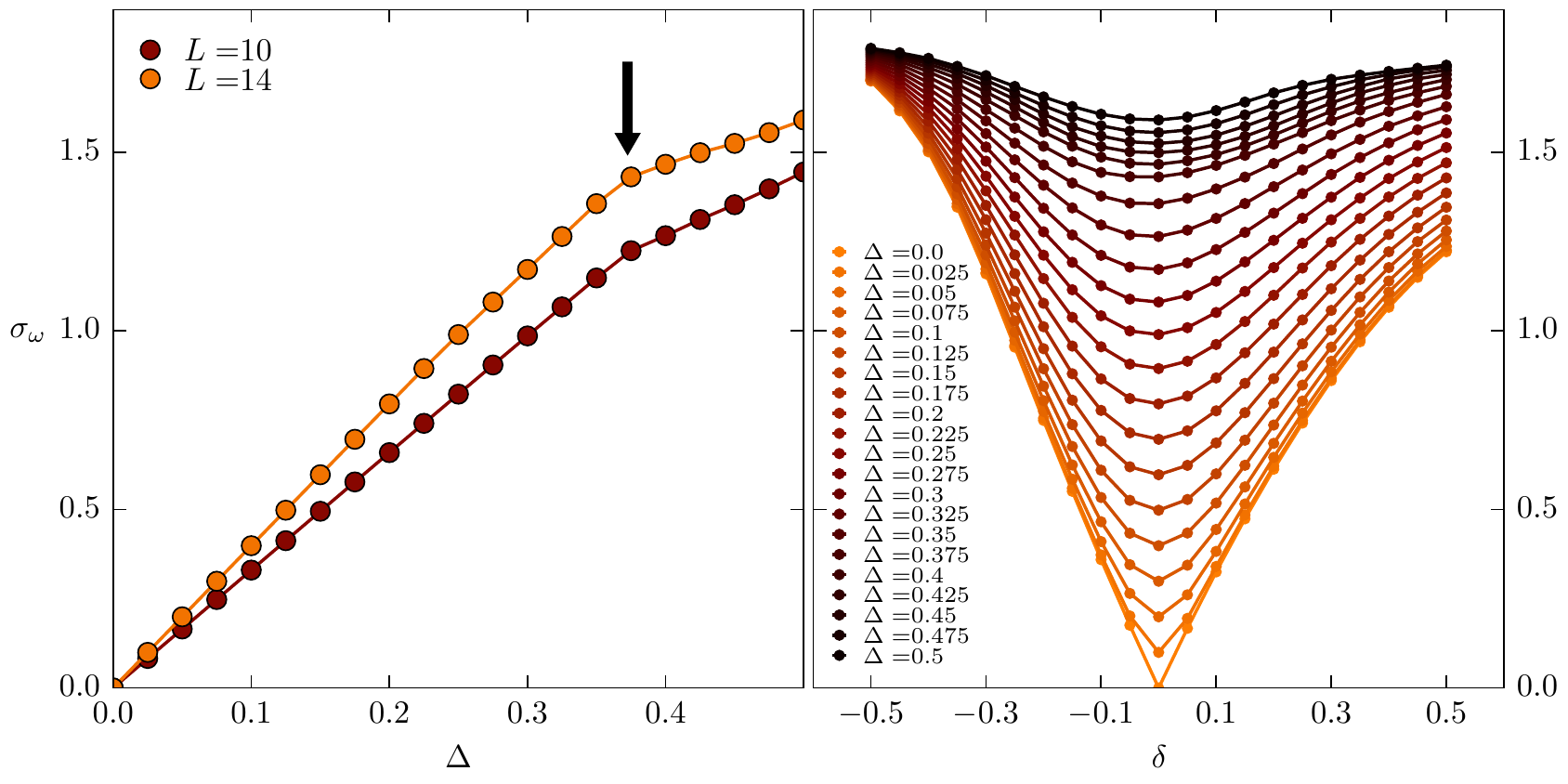}
\par\end{centering}
\caption{\label{fig:broadening} Left: Standard deviation $\sigma_{\omega}$
    of quasienergies versus interaction strength, $\Delta$ for $\delta=0$ (see (\ref{eq:detuning-definition})).
 Right: Standard deviation of quasienergies as a function of detunings $\delta,$ for various
interaction strengths, $\Delta$. The arrow indicates the
interaction strength at which the quasienergy spectrum starts to wrap
in the Floquet Brillouin zone, essentially at the same interaction
strength for $L=12$ and $L=14.$}
\end{figure}

Finally we examine the broadening of the Floquet spectrum upon addition
of interactions. As explained in Sec.~\ref{sec:Dynamical-localization},
the spectrum of the operator $\hat{U}\left(T,0\right)$ in (\ref{eq:definition-U})
collapses to a single point at the noninteracting ($\Delta=0$) dynamically
localized point $\delta=0$, resulting in a trivial stroboscopic dynamics.
A perturbative reasoning suggests that a finite interaction will lift this
massive degeneracy and broaden the quasienergy spectrum. We have confirmed this
expectation by evaluating the quasienergies $\epsilon_{\alpha}$,
defined as $\hat{U}\left(T,0\right)\ket{\alpha}=\exp\left(-\I\epsilon_{\alpha}T\right)\ket{\alpha}$,
by exactly diagonalizing $\hat{U}\left(T,0\right)$ up to systems
sizes $L=14.$ The width of spectrum was calculated from the standard
deviation $\sigma_{\omega}$ of the $\epsilon_{\alpha}$ (after the
quasienergies were shifted to the center of the band to avoid spurious
widening due to wrapping around the Floquet Brillouin zone). The results
are shown in Fig.~\ref{fig:broadening}, where we see that the spectrum
broadens upon addition of interactions, however the location of the
minimal (but still finite) width remains at zero detuning,
$\delta=0$ (see (\ref{eq:detuning-definition})). This is consistent with
our observation that slowest transport occurs at this point (Fig.~\ref{fig:MSD-deriv-detun}).

\section{Discussion\label{sec:Discussion}}

In this work we have examined the stability of dynamical localization
to the addition of interparticle interactions. We have found that
while dynamical localization is destroyed by interactions, transport
is significantly suppressed when the interacting system is at the
noninteracting dynamically localized point (see right panel of Fig.~\ref{fig:broadening}
where the minimum of the width occurs at zero detuning, $\delta=0$ for
all considered interaction strengths $\Delta$). The slow transport
allows us to find clear evidence of diffusive transport with a diffusion
constant proportional to the interaction strength, indicating that
scattering is dominated by interparticle collisions. We have also
studied the dependence of the width of the quasienergy spectrum on
the interaction strength. While the spectrum is completely degenerate
for zero interactions, leading to trivial stroboscopic dynamics, for
finite interaction the degeneracy is lifted and the width appears
to be proportional to $\Delta$ before it saturates to its maximum
value fixed by the driving frequency. From Fig.~\ref{fig:broadening}
one can see that for a \emph{fixed} interaction strength the width
increases with system size. This leads us to speculate that in the
thermodynamic limit an arbitrarily weak interaction causes the spectrum
to occupy the full allowed quasienergy range. While it is tempting
to conclude that the width of the spectrum corresponds to the rate
with which the dynamics unfolds, this is not true in general. For
example, a finite width occurs for Floquet-MBL systems while the dynamics
is frozen \cite{Lazarides2014,Ponte2014}. We therefore can
only conclude that when the spectrum broadens dynamical localization
may be destroyed, but the precise nature of the underlying dynamics has to
be assessed by other means (as we have done in Figs.~\ref{fig:MSD-deriv}
and \ref{fig:MSD-deriv-detun}).

One interesting question on which our work might have some bearing
is the locality of the effective Hamiltonian, $\hat{H}_{F}$. This
quantity is defined from $\hat{U}\left(T,0\right)$ in (\ref{eq:definition-U})
as $\hat{U}(T,0)=\exp\left(-\I\hat{H}_{F}T\right)$ and is the generator
of the stroboscopic time evolution. It has been speculated that for
generic Floquet systems which are not MBL the effective Hamiltonian
$\hat{H}_{F}$ is nonlocal \cite{DAlessio2013,DAlessio2014}. Since
it is generally believed that spreading of correlations in systems
with nonlocal \emph{static} Hamiltonians will be superballistic, \cite{Gong2014,Richerme2014,Foss-Feig2015,Maghrebi2016}
one may wonder whether the diffusive transport we observe in this
work points toward a \emph{local} effective Hamiltonian. We would
like to argue that due to the ambiguity in the definition of the effective
Hamiltonian the question is not well posed. The ambiguity follows
from the fact that $\hat{U}\left(T,0\right)$ and therefore all stroboscopically
measured observables are invariant under the transformation $\hat{H}_{F}'=\hat{H}_{F}+\omega\sum_{\alpha}n_{\alpha}\ket{\alpha}\bra{\alpha}$,
where $\ket{\alpha}$ are the eigenstates of $\hat{U}\left(T,0\right)$
and $n_{\alpha}\in\mathbb{Z}$. Since in general the projectors $\ket{\alpha}\bra{\alpha}$
are nonlocal objects this renders the locality of $\hat{H}_{F}$ a
property which cannot be inferred from physical observables. While we believe that
the discussion of locality of $\hat{H}_{F}$ is irrelevant as far
as physical observables are concerned, it remains an interesting
question whether the ultimate fate of driven interacting systems is
encoded in the locality of the corresponding \emph{Floquet operator,}
$\hat{U}\left(T,0\right)$, which \emph{is} a physical quantity and
whose locality can be assessed, for example using the operator entanglement
entropy \cite{Zhou2017}. We leave this question for future work. 

\section{Acknowledgments}
YB acknowledges funding from the Simons Foundation (\#454951, David
R. Reichman). Our code uses the libraries PETSc \cite{petsc-efficient,petsc-user-ref,petsc-web-page}
and SLEPc \cite{hernandez_slepc:_2005}. This work used the Extreme
Science and Engineering Discovery Environment (XSEDE), which is supported
by National Science Foundation grant number ACI-1548562. This project
has received funding from the European Union's Horizon 2020 research
and innovation programme under the Marie Sk\l odowska-Curie grant
agreement No. 747914 (QMBDyn). DJL acknowledges PRACE for awarding
access to HLRS's Hazel Hen computer based in Stuttgart, Germany under
grant number 2016153659.



\bibliography{lib_yevgeny,lib_tmp_acl,lib_david}

\nolinenumbers

\end{document}